\documentclass[usenatbib]{emulateapj}
\usepackage{epsfig}
\usepackage{amsmath}
\usepackage{amssymb}
\catcode`\@=11
\def\gta{\ifmmode{\,\mathrel{\mathpalette\@versim>\,}}
    \else{$\,\mathrel{\mathpalette\@versim>}\,$}\fi}
\def\lta{\ifmmode{\,\mathrel{\mathpalette\@versim<\,}}
    \else{$\,\mathrel{\mathpalette\@versim<}\,$}\fi}
\def\@versim#1#2{\lower 2.9truept \vbox{\baselineskip 0pt \lineskip
    0.5truept \ialign{$\m@th#1\hfil##\hfil$\crcr#2\crcr\sim\crcr}}}
\catcode`\@=12  
{\newif\ifnotend
\notendtrue
\def\veclist{ABCDEFGHIJKLMNOPQRSTUVWXYZabcdefghijklmnopqrstuvwxyz.}
\def\top#1#2.{#1}
\def\tail#1#2.{#2.}
\loop\expandafter\xdef\csname v\expandafter\top\veclist\endcsname%
{{\noexpand\bf\expandafter\top\veclist}}
\edef\veclist{\expandafter\tail\veclist}
\if\veclist.\notendfalse\fi\ifnotend\repeat}

\def\kms{\,{\rm km}\,{\rm s}^{-1}}

\def\Gyr{\,{\rm Gyr}}
 
\def\pc{\,{\rm pc}}
\def\kpc{\,{\rm kpc}}

\def\metr{{\rm m}}

\def\figref#1{Fig.~\ref{#1}}

\def\ovcv0{\overline{V_0}}

\newcommand {\GHz}{{\rm GHz}}

\newcommand {\Rsun}{{R_{0}}}

\newcommand {\Usun}{{U_{\!\odot}}}

\newcommand {\Wsun}{{W_{\!\odot}}}
\newcommand {\Vsun}{{V_{\!\odot}}}
\newcommand {\vlos}{{v_{\rm los}}}
\newcommand {\vgal}{{v_{\rm gal}}}
 
\newcommand{\beq}{\begin{equation}}
\newcommand{\eeq}{\end{equation}}

\shorttitle{Kinematic Detection of the Nuclear Disc}
\shortauthors{Sch\"onrich, Aumer \& Sale}

\begin{document}

\title{Kinematic Detection of the Galactic Nuclear Disc}

\author{Ralph Sch\"onrich, Michael Aumer, Stuart E. Sale}
\affil{Rudolf Peierls Centre for Theoretical Physics, 1 Keble Road, Oxford, OX1 3NP, UK}
\email{Email: ralph.schoenrich@physics.ox.ac.uk}


\label{firstpage}

\begin{abstract}
We report the detection of the Galactic nuclear disc in line-of-sight kinematics of stars, measured with infrared spectroscopy from APOGEE. This stellar component of the nuclear disc has an extent and rotation velocity $V\sim120\kms$ comparable to the gas disc in the central molecular zone. The current data suggest that this disc is kinematically cool and has a small vertical extent of order $50\pc$. The stellar kinematics suggest a truncation radius/steep decline of the stellar disc at a galactocentric radius $R\sim150\pc$, and provide tentative evidence for an overdensity at the position of the ring found in the molecular gas disc.
\end{abstract}


\section{Introduction}

Through most of modern astronomy, empirical studies of the central regions of the Milky Way focused on radio observations of the central gas \citep[cf.][]{Binney91,SormaniI}. Visible photometry and spectroscopy are nearly impossible due to the extinction from the Galactic mid-plane and the central regions themselves. Even Baade's Window is at a latitude $|b| \sim 3$ degrees \citep[cf.][]{Gonzales12}. While radio observations have revealed the a $\sim200 \pc$ gas disc in the central molecular zone (CMZ) of the Milky Way and its embedment in the Galactic bar both in density and gas kinematics \citep[][]{Rougoor60,Peters74}, stellar evidence is scarce. Apart from sporadic studies on stellar kinematics via OH masers \citep[see e.g.][]{Habing83,Lindqvist92,Deguchi04}, who found systematic rotation in the near-centre MASERS, the nuclear disc has been indirectly inferred by modelling the photometric observations with estimated density profiles \citep[][]{Catchpole90,Launhardt02}.

The advance of infrared spectroscopy has now led to large systematic surveys of the Galactic mid-plane, including the Galactic centre, like the Apache Point Galactic Evolution Experiment (APOGEE, Majewski et al. in prep.). Earlier studies mostly focused on the detection of stars around the central black hole \citep[e.g.][]{Schoedel02} and using those orbits to constrain the black hole parameters and distance \citep[see][]{Ghez09,Gillessen09}. Note that the nuclear disc studied here is a factor of $30$ larger and likely of a different origin than the $\sim 5 \pc$ disc(s) around the central black hole \citep[][]{Paumard06}. Studies of the broader nuclear region of the Milky Way \citep[see e.g.][]{Schultheis03,Cunha07} were mostly limited to rather small sample sizes or single clusters. E.g. \cite{Matsunaga15} claim that the kinematics of their $4$ Cepheid stars are consistent with the Nuclear Disc.

The CMZ/nuclear gas disc is a direct consequence of the Galactic bar. Gas flows inward along the leading edge of the bar. Although some star formation has been detected along the leading edges of extragalactic bars \citep[][]{Sheth02}, most of this gas will eventually cross from a neighbouring $x_1$-orbit onto the $x_2$-disc \citep[][]{Albada82}. There it will accumulate until it reaches densities sufficient for star formation. Hence, some of the gas in the nuclear disc forms stars, a minor fraction is accreted onto the central black hole, and a major part will be expelled by stellar and Active Galactic Nucleus (AGN) feedback \citep[cf. the bipolar outflow from the centre observed by][]{BH03}. Although this central star formation has been studied in several papers \citep[e.g.][]{Serabyn96,Loon03,Figer04,Yusuf09,Kruijssen14}, the nuclear stellar disc has not been directly detected from stellar spectra. Here we report the kinematic detection of the stellar nuclear disc using a sample of stars from the APOGEE survey. 

Data-related information is found in the next section. Section 3 presents observational evidence, a comparison to gas motions, and a qualitative exploration with toy models. We conclude in Section 4.

\begin{figure}
\epsfig{file=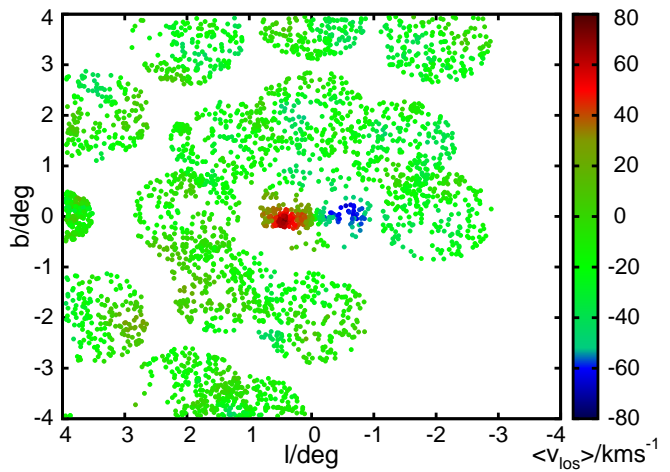,angle=-90,width=\hsize}
\caption[]{Overview of APOGEE stars (coloured dots) near the Galactic centre in Galactic longitude $l$ and latitude $b$. Colours represent the mean line-of-sight velocity $\vlos$ of each star and its closest $29$ neighbours. Note the division into plates/fields, and the clear dipole structure in $\vlos$ around the Galactic centre.
}\label{fig:geo}
\end{figure}

\newpage
\section{Data}\label{sec:data}
We use stellar positions and line-of-sight velocities from APOGEE, downloaded from the Casjobs system of the Sloan Digital Sky Survey III \citep[SDSSIII,][]{SDSSIII}. APOGEE is the first large infrared spectroscopic survey of the Milky Way. Its multi-object fibre spectrograph simultaneously obtains about $\sim300$ $H$-band spectra with resolution $R\sim22500$ per plate/exposure (see the plate structure in Fig. 1). The survey selection function \citep[][]{APtarget} aims at giant stars with the lowest possible complexity and bias. For each plate, all stars with (de-reddened) colour $(J-K_S)_0>0.5$ and de-reddened $H_0$-band magnitude within $6<H_0<11$ (for the central MW fields) are selected with equal probability. As discussed in \cite{Aumer15}, this selection favours young stellar populations (comparing ages of $1$ and $10\Gyr$) by about an order of magnitude, since the larger relative lifetime in evolved stages of massive stars dominates over the opposite effect from the initial mass function. 

While the APOGEE selection function \citep[][]{APtarget} was designed to be unbiased, the Galactic Centre plate strongly favours positive Galactic longitudes. This asymmetry is not physical, but is caused by the survey for two reasons: manual selection and photometric crowding. 

The Galactic Centre field in APOGEE \citep[termed "GALCEN" in][]{APtarget} contains $233$ out of $379$ stars that were added to the survey outside the normal selection criteria for comparison with pre-existing spectroscopy. The good news is that neither APOGEE's secondary sample, nor the pre-existing samples were kinematically selected. Hence, the detection of the disc kinematics is fine, while the relative foreground contamination and relative numbers within the structure may be distorted.

The APOGEE target selection uses 2MASS photometry \citep[][]{2MASS} combined with Rayleigh Jeans Colour Excess \citep{Majewski11} reddening estimates based on Spitzer-GLIMPSE data \citep[][]{GLIMPSEI,GLIMPSEII}. The 2MASS photometry in this field is compromised by inhomogeneous extinction and crowding: the smaller extinction at negative longitudes, which is connected to the asymmetry of the CMZ \citep[CMZ,][]{Bally88}, and hence stronger crowding reduces the number of target candidates passing the quality criteria. This prevents a clean assessment of the detailed structure of the nuclear disc.

APOGEE does not offer stellar parameters for most objects in the Galactic centre. To diminish foreground contamination, we thus have to rely on photometric methods. Obviously, the handful of telluric standards can be removed by imposing $(J-K)_0>0.5$. We obtain a cleaner subsample via the lower extinction of foreground stars, requiring that stars have a valid reddening estimate $A_K>3$ based on WISE photometry \citep[cf.][]{Blum96,Figer04,Gonzales12}. This may preferentially remove stars on the front edge of the nuclear disc, a bias which we assess by comparison of the two subsamples.

To compare the stellar kinematics with the gas in the CMZ, we use publicly available data cubes from the AST/RO survey \citep[Antarctic  Submillimeter Telescope and Remote Observatory,][]{ASTRO}. For the longitude-velocity ($lv$) plot, we sum the emission temperatures for all $73$ latitude bins within the survey's latitude range of $-0.3<b/\deg<0.2$ from the CO $J=4\to3$ lines at $461.041\GHz$. While this survey has a mildly lower resolution compared to later data compilations \citep[see e.g.][]{Jones12}, it offers the advantage of publicly available and well documented data, while it covers the main features observed in the $lv$-plane by other surveys.

We translate line-of-sight velocities $\vlos$ into Galactic rest frame velocities $\vgal$, defined as
\begin{equation}
\vgal=\vlos+\Usun\cos(l)\cos(b)+\Vsun\sin(l)\cos(b)+\Wsun\sin(b)\text{.}
\end{equation}
For the solar galactocentric distance we assume  $\Rsun=8.27\kpc$ and for its motion against the Galactic rest frame $(\Usun,\Vsun,\Wsun)=(14,250,7)\kms$ \citep{Gillessen09,McMillan11,S12,SBD10}.

\section{Kinematic detection of the nuclear disc}

The rotation of the nuclear stellar disc of the Milky Way is evident from \figref{fig:geo}, limited to the innermost degree around the Galactic centre. Stars at negative longitudes approach with mean $\vlos\sim-80\kms$ and recede with the same speed at positive longitudes. This pattern is limited to latitudes $|b|\lesssim0.4\deg$ and is hence directly connected to the disky overdensity identified e.g. by \cite{Launhardt02}, who find a disc radius of $\sim230\pc$ and a scale height of $\sim45\pc$ (though the bulk of their $60\mu\metr$ emission is within $1$ degree of the Galactic centre). 

The case that this is indeed a rotating nuclear disc is simple: Despite the foreground contamination, the average $\vlos$ shows a shift of more than $100\kms$ on this small scale. The feature is far too dominant for any background effect behind the Galactic centre. Spiral structure may shape velocities by $\sim 10 \kms$, an order of magnitude less than this observation. Also, its length scale would by far exceed the $<100\pc$ implied by the angular size. The other candidate would be a connection to the velocity features discovered by \cite{Nidever12}. However, these are again too weak in numbers and too extended covering an angular scale of ${\Delta}l>10$ degrees. Explaining a mean velocity shift centered around a mean $<\vlos>\sim0$ would require that miraculously the foreground feature is observed exclusively on one side and the background feature on the other side of the Galactic Centre. Last, no systematics are detected in the motion above $|b|\sim0.4\deg$, i.e. altitudes larger than $\sim50\pc$, confirming a disc roughly within the Galactic plane. 
Hence, the only possible explanation is a rotating nuclear disc around the Galactic Centre.

\begin{figure}
\epsfig{file=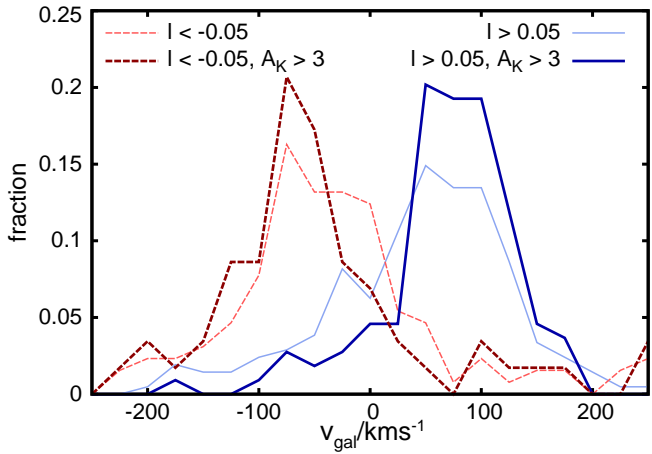,angle=-90,width=\hsize}
\caption[]{Normalised velocity distributions of APOGEE stars with $|l|,|b|<1\deg$. The sample was split into stars with $l>0.05\deg$ ($208$ stars) and $l<-0.05\deg$ ($129$ stars) to present the two halves of the central disc. Even without any further selection, the two sides of the nuclear disc dominate the sample. The foreground contamination is reduced when we narrow the selection to highly reddened stars ($A_K>3$, $109$ and $58$ objects).
}\label{fig:vdis}
\end{figure}

In \figref{fig:vdis} we show the velocity distributions for stars with $|l|,|b|<1\deg$, dividing the field into stars with $l>0.05\deg$ and $l<-0.05\deg$. Already the entire sample is dominated by the strong odd part in the velocity distribution coming from the two sides of the disc at velocities around $50-120\kms$. The small velocities are dominated by foreground stars in the disc mid-plane, while both sides show a couple of high velocity stars, which are associated with the bar/bulge region \citep[stellar populations streaming alongside the bar have tails up to $|\vgal|\sim300\kms$, see][]{Nidever12,Aumer15}. The disc kinematics are cleaned up via the extinction cut $A_K>3$ (thicker lines) .

\begin{figure*}
\epsfig{file=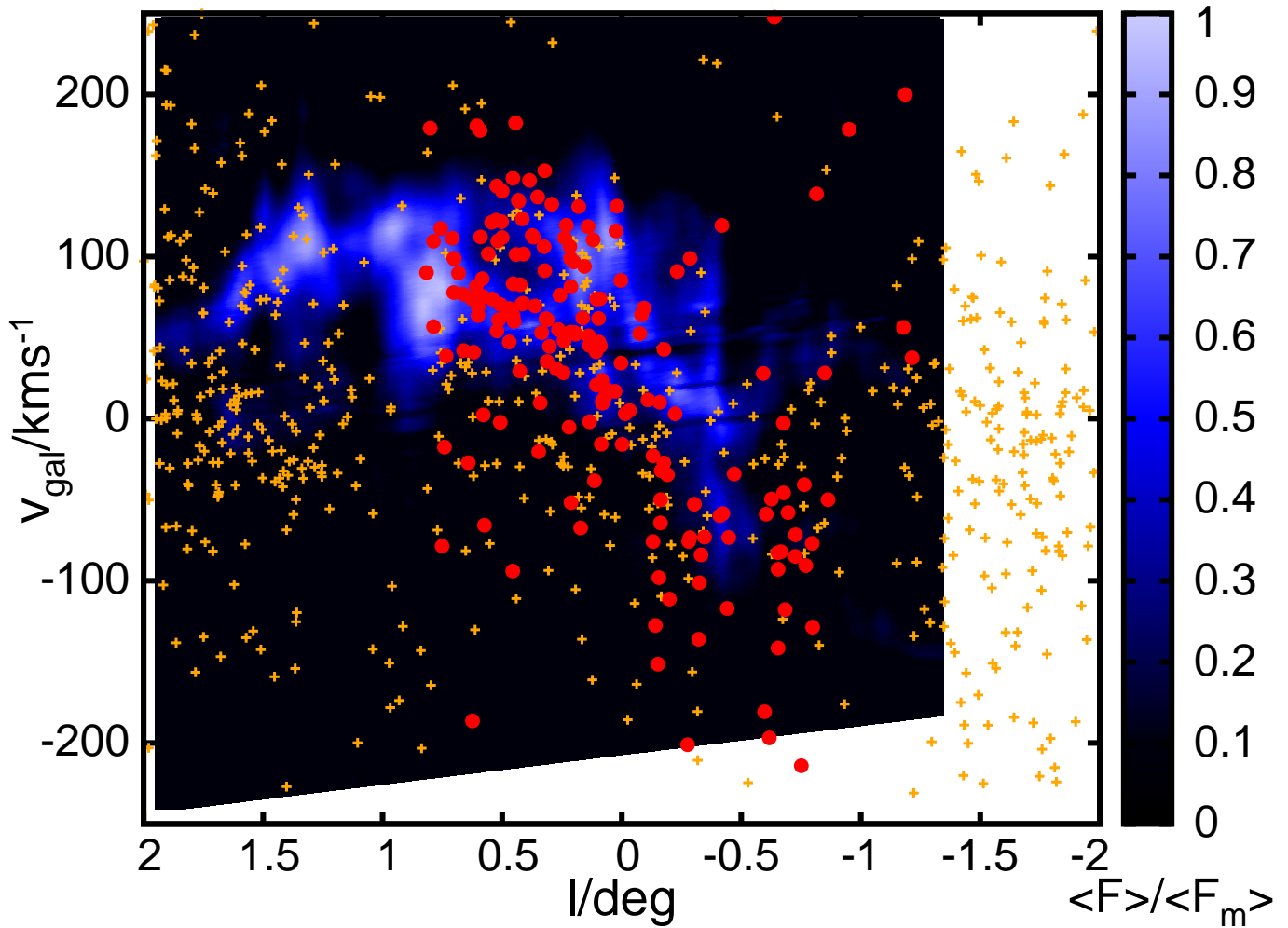,angle=-90,width=\hsize}
\caption[]{Latitude-velocity ($lv$) plot of stars and gas in the Galactic centre region. The colour code gives the intensity of CO $J=4\to3$ emission from the AST/RO survey integrated over all $|b|<0.3\deg$ in units of the maximum emission density. The $\vlos$ range not covered by the survey is left blank, the cut-off has a slope due to the translation into $\vgal$. Orange crosses indicate APOGEE stars at latitudes $|b|<1\deg$, red bullets the likely disc members with estimated $K$-band extinction $A_K>3$.
}\label{fig:scatt}
\end{figure*}

We now compare the stellar kinematics in the $lv$-plane to the motions of the molecular gas in the known nuclear gas disc. In \figref{fig:scatt} we display the integrated emission in the CO $J=4\to3$ transition with $|b|<0.3\deg$ in $\vgal$ (y-axis) versus longitude $l$ with blue to white shades. We can clearly see the ridge of emission around $115\kms$ at positive longitudes, as well as a lower, sloping ridge connecting the bright emission of the Sgr B2 region at $(l,\vgal)\sim(0.7\deg,70\kms)$ to the origin. The latter has been interpreted as a ring-shaped region of enhanced radio emissions at a radius of $R\sim150\pc$ around the Galactic centre \citep[][]{Molinari11}. We overplot APOGEE stars with $|l|,|b|<1\deg$ (orange crosses). To curb foreground contamination again, we mark the high-reddening stars with $A_K>3$ as red bullets. While the velocity distribution of low-reddening stars is nearly independent of longitude, consistent with their being foreground stars, the high-reddening stars display the point symmetry typical for disc kinematics, plus moderate contamination. They also show some tentative structure along the two ridges defined by the molecular gas. 

\begin{figure}
\epsfig{file=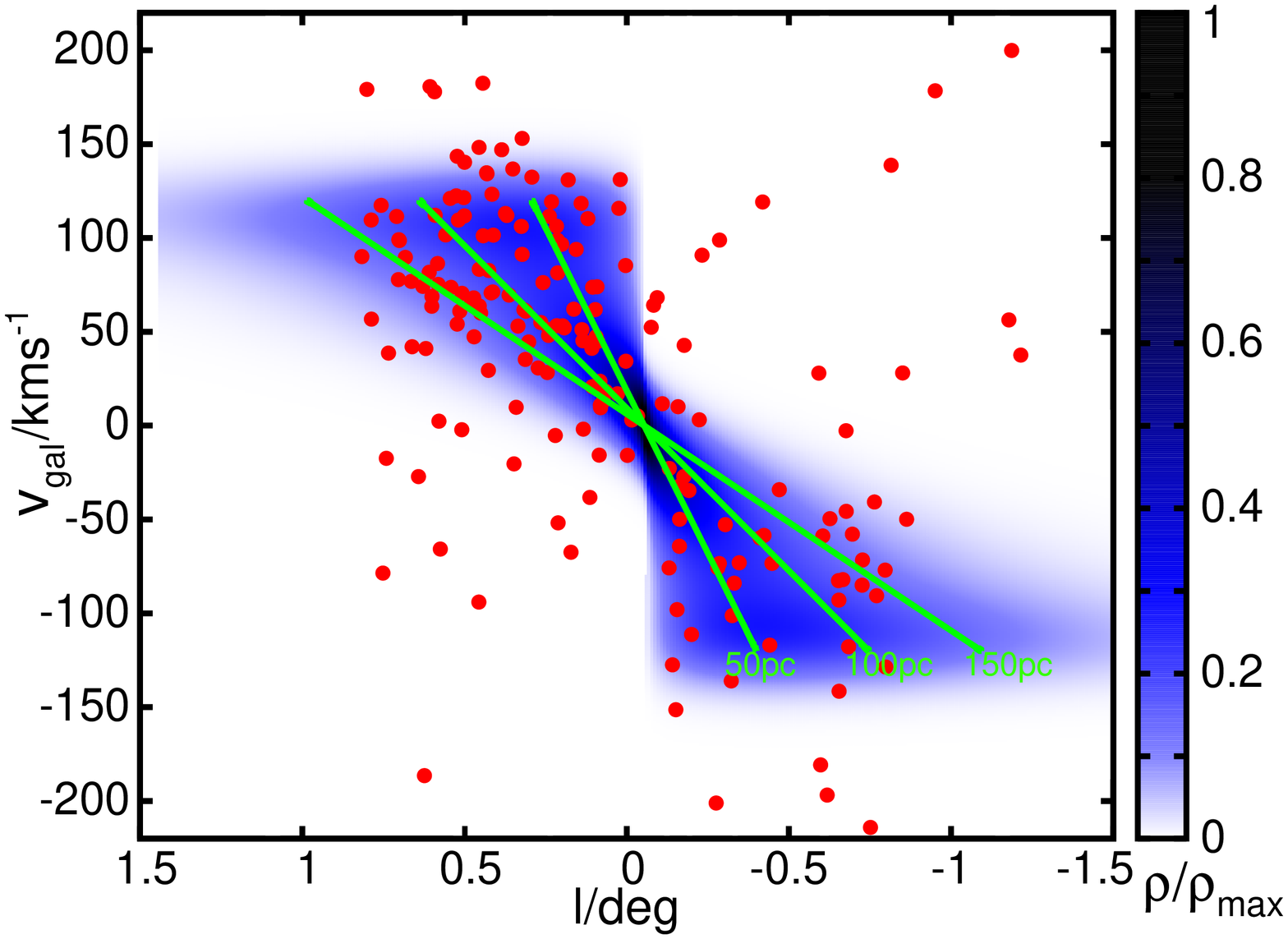,angle=-90,width=0.98\hsize}
\epsfig{file=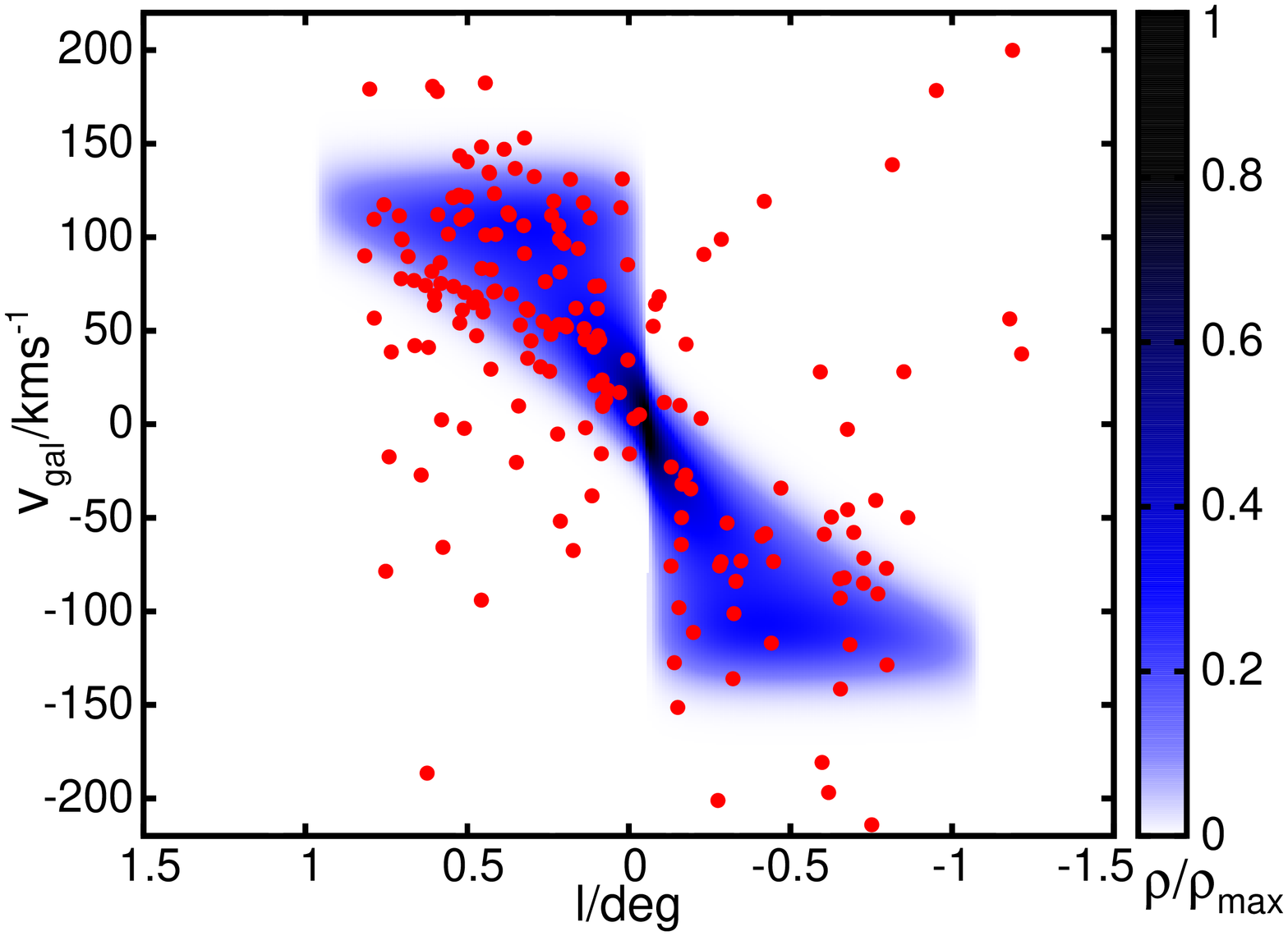,angle=-90,width=0.98\hsize}
\epsfig{file=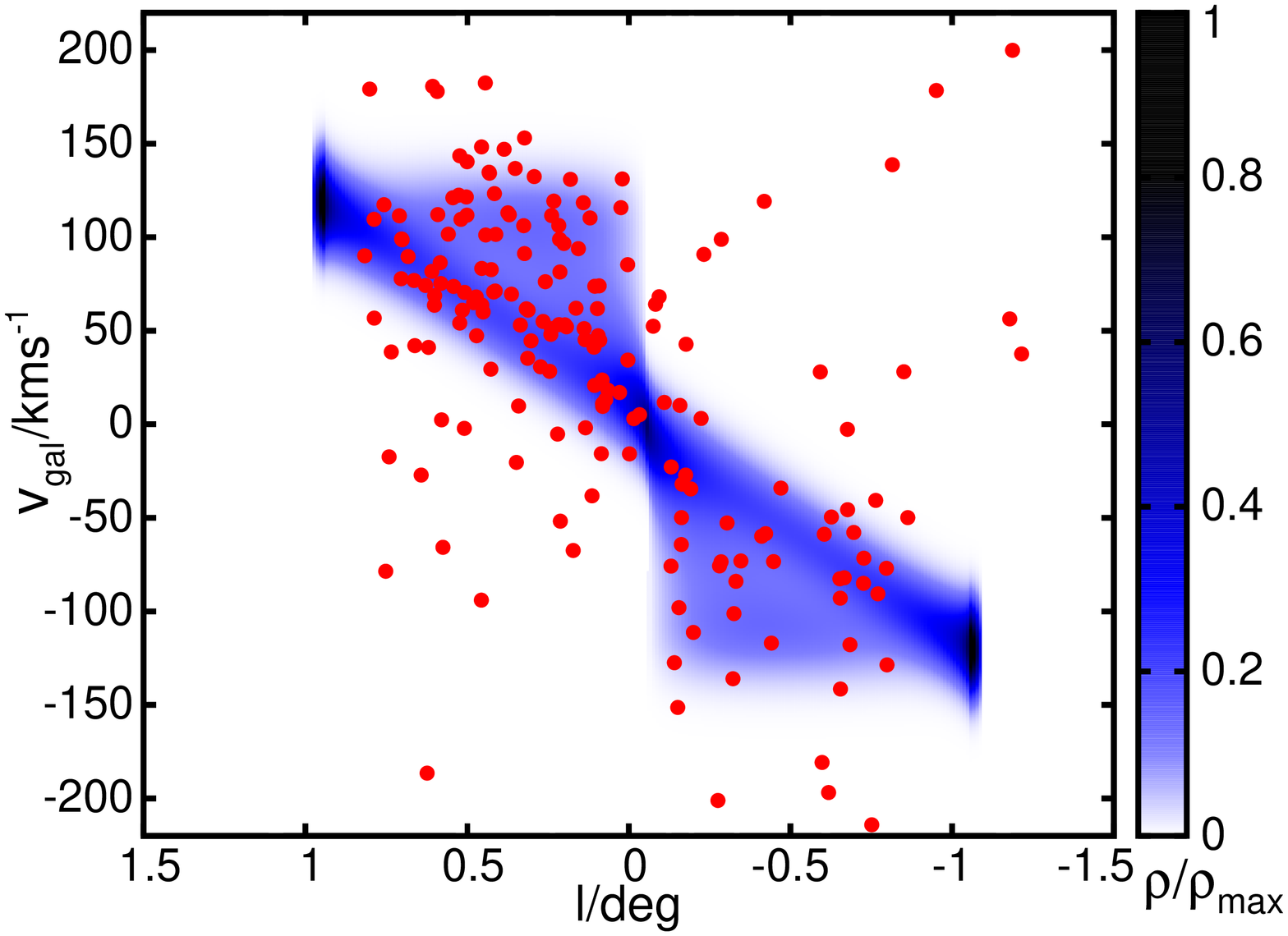,angle=-90,width=0.98\hsize}
\caption[]{Latitude-velocity plot of high-reddening APOGEE stars from \figref{fig:scatt} at $|b|<0.4\deg$ (red circles) compared with three toy models (blue shading) with increasing complexity: A simple exponential disc (top), an exponential disc with a truncation radius beyond which the density drops steeply (middle), and a denser ring near the truncation radius (bottom, parameters from Table $1$ for $A_K>3$). In the top panel we add as green lines the line-of-sight velocities that perfect rings with rotational speed $V=120\kms$ and radii $50$, $100$, and $150\pc$ would have. Note that there are almost no APOGEE stars beyond $|l|>0.7\deg$ due to the survey geometry.
}\label{fig:mod}
\end{figure}

Detailed quantitative exploration with consistent kinematic models has to await a decent 3D-reddening map and reasonable stellar parameters to assess the foreground contamination. However, for a qualitative discussion we provide three mock models in \figref{fig:mod}. The plotted mock models are axisymmetric with a constant azimuthal streaming velocity $V_\phi\sim120\kms$. We shift the longitude of the toy model centre by $\Delta l=-0.054\deg$ to the position of $Sgr A^*$. The data favour this, as the sign change in mean $\vgal$ occurs at negative longitude.

\begin{table}
\caption{\rm Fit parameters from equation(2) for stars with $|b|<0.4\deg$. Errors are given by $\Delta\ln(P)=1/\sqrt{2}$ with full variation of the other parameters.}\label{tab:zfitpar}
\begin{tabular}{|l|c|c|c|}\hline
name&parameter&all&$A_K>3.0$\\\hline
{\rule{0pt}{2.6ex}}azimuthal vel.&$V_\phi/\kms$ & $123.0_{-7}^{+19}$&$120_{-9}^{+14}$\\
{\rule{0pt}{2.6ex}}scale length&$R_d/\pc$&$52_{-17}^{+51}$&$60_{-22}^{+70}$\\
{\rule{0pt}{2.6ex}}radius of ring&$R_r/\pc$&$150_{-20}^{+22}$&$147_{-16}^{+16}$\\
{\rule{0pt}{2.6ex}}ring/centr. density&$f_r$&$1.6_{-1.2}^{+1.5}$&$3.0_{-1.7}^{+2.9}$\\
{\rule{0pt}{2.6ex}}trunc. radius&$R_t/\pc$&$150_{-16}^{+\inf}$&$147_{-14}^{+\inf}$\\
{\rule{0pt}{2.6ex}}contam. fraction&$f_c$&$0.135_{-0.043}^{+0.055}$&$0.059_{-0.021}^{+0.032}$\\
{\rule{0pt}{2.6ex}}contam. vel. disp.&$\sigma_c/\kms$&$60_{-11}^{+19}$&$54_{-13}^{+29}$\\
{\rule{0pt}{2.6ex}}log likelihood&$\ln(P)$&$-440.2$&$-325.4$\\\hline
\end{tabular}
\end{table}

The predicted velocity distributions are folded with a Gaussian kernel with $\sigma=15\kms$ to mimic the random motion of stars (which in reality should be neither isotropic, nor constant in $R$). The dispersion has an upper limit by the sharpness of the sloping density ridge, and a lower limit by the expectation for secular heating. Model complexity increases from top to bottom, starting from a simple exponential disc. The middle panel introduces a truncation radius, and the bottom model adds a ring of increased density near the truncation. The toy model for the stellar surface density reads:
\begin{equation}
\Sigma_d(R)=e^{-R/R_d}\cdot\begin{cases}
1&\text{if$\,R{\le}R_t$} \\
e^{-(R-R_t)/R_{d,2}}&\text{if$\,R>R_t$},
\end{cases}
\end{equation}
with cylindrical radius $R$, scale-length $R_d$, truncation radius $R_t$. We set the drop-off steepness $R_{d,2}=1\pc$. To the bottom panel we add a $5\pc$ wide, constant surface density ring, centred on a radius $R_r$ and with density $f_r$ relative to the disc centre. The model without truncation has $R_{d,2}\to\infty$.

The mild density maximum near the circular velocity derives from the comparatively large distance range within which the azimuthal direction of the disc is nearly collinear with the line of sight. This is enhanced by the density profile, as the tangential parts probe the smallest radii. A radial truncation removes at fixed $l$ the stars at small velocities, and limits the extent of the disc in longitude. Note that the longitude limits of the APOGEE field do not affect our fits since we separately normalise narrow bins in longitude. A ring with enhanced density (see bottom panel) produces a straight density ridge between the origin and the tangential longitude of the ring, where it ends in a pronounced density maximum.

Fit parameters for the full mock model are listed in Table\ref{tab:zfitpar}. In these fits we demand that $R_t < R_r$ and allow for a naive Gaussian foreground contamination with dispersion $\sigma_c$ and a constant (in $l$) contamination fraction $f_c$. Fits are restricted to positive $l$, sampled into $0.1$ degree wide longitude bins with separate normalization in star count. Negative longitudes are excluded due to the more uncertain selection function and anyway low number counts. We limit the fitting range to $|\vgal|<130\kms$. This curbs contamination from high velocity structures, i.e. fat tails in the contaminating velocity distributions \citep[cf.][]{Aumer15}, while still covering some of the decline in disc density above the rotation speed. Comparison between the results for samples with and without the reddening cut $A_K>3.0$ confirms that this cut removes most of the foreground contamination, while the fit parameters stay mostly unaffected. The data suggest a truncation radius, though its removal can, within the significance limits, be compensated by a shorter scale length ($<50\pc$). If we drop the condition $R_t<R_r$ the truncation radius moves inwards to $80\pc$ with a significant improvement of the likelihood. The fits demand a ring with comparable densities to the disc centre (uncertain due to unknown width and dispersion), and with a radius around $150\pc$. 
Despite this seemingly robust result, we caution that these mock models are very unsatisfactory and should no be considered more than a first guidance. The truncation radius is in a zone of heavy differential effects in reddening, and the APOGEE selection is not well constrained.

These toy models underline the need for additional data: In particular the bottom panel of \figref{fig:mod} shows the need for a larger Galactic centre sample covering the suspected tips of the nuclear disc in longitude, which would also be important to assess a possible non-circularity.

\newpage
\section{Conclusions}\label{sec:p7conclude}

We report the detection of the Galactic nuclear disc from stellar spectra of the APOGEE survey. 

Even without removing any foreground contamination, a dipole in line-of-sight velocities is prominent in the data with a velocity shift $\gtrsim100\kms$, limited to within $\sim1$ degree from the Galactic Centre. The only viable explanation for such a strong mean velocity contrast on smallest scales is a rotating disc in the central $\sim(150-180)\pc$.

We show that the data are consistent with a nearly axisymmetric distribution with an azimuthal speed of $\sim120\kms$. This is in line with gas motions of the CMZ derived from radio observations. Signatures of rotation in the OH/IR objects were found by \cite{Lindqvist92}. The disc appears confined to latitudes $|b|\lesssim0.4\deg$, i.e. an altitude $|z|\lesssim50\pc$ consistent with the photometric density estimates from \cite{Launhardt02}. Due to the selection uncertainties and spatial limits, we do not use the spatial densities. Consequently the limited plate extent with $|l|\lesssim0.7\deg$ does not alter our conclusions, but only limits our stated significance. The stellar kinematics alone provide evidence for a ring of enhanced density around $R\sim150\pc$, outside which the data hint to a truncation/steeper decline. This ring feature coincides with the molecular ring found in the gas motions. Its small width in line-of-sight velocities suggests a kinematically cool population. The bulk of the nuclear disc mass resides within $\sim150\pc$, with indications for a steeper decline outside the ring. Not allowing for a truncation, the scale length would drop to $R_d\lesssim50\pc$ with nearly the same effect.
This implies a somewhat smaller radial scale than the $\sim230\pc$ found in photometry by \cite{Launhardt02}, though it is in line with the bulk of their $60\mu\metr$ emission.

While the detection of the nuclear disc is beyond the tightest significance limits (without any clean-up the line-of-sight velocity signal from \figref{fig:geo} exceeds $10\sigma$), its precise nature is more ambiguous. We explore the possible shape with toy models. However, any quantitative approach has to await the creation of 3D-reddening maps, and a manual analysis of the stellar parameters, to assess the precise foreground contamination and the selection function of stars in this region. On the other hand, this finding facilitates analysis of nuclear disc stars, as they can be clearly identified by their kinematics.

These uncertainties prevent an assessment of possible non-circularity in the nuclear disc. A mild non-circularity is expected, as this disc is born from gas on orbits of the $x_2$ family and affected by the rotating bar \citep[][]{Binney91}, some might derive from possible spiral arms within this disc. \cite{Molinari11} suggest an even more complex shape, though the kinematic stability of such a structure is doubtful; improvements are discussed in \cite{Kruijssen15}. 

In a recent paper, \cite{Debattista15} claimed to have identified a $\sim1\kpc$ nuclear disc based on the high-velocity feature of \cite{Nidever12} at longitudes $l\sim10$ degrees. Although the idea is intriguing, they have not presented conclusive evidence that their model can fit the observed $\vlos$ distributions. The simulations of \cite{SormaniI} suggest such a large $x_2$ disc is unlikely. Also, it would require significant flattening of the density profile suggested both by this study and by \cite{Launhardt02}. More critically, these observations provide no proof/detection of a $\sim1\kpc$ sized $x_2$ disc, since \cite{Aumer15} have identified them with orbits alongside the bar (largely $x_1$ orbits), in a model that provides a good fit to the observations. A similar conclusion was drawn by \cite{Molloy15}. We would like to stress again that those high velocity kinematic features cannot explain the observations in the central degree: their stars are expected to have larger velocities ($\sim200\kms$) than the nuclear disc, which is $\sim100\kms$ slower. More importantly they would be found on both sides of the Galactic centre, i.e. both the background stars at $\sim+200\kms$ and the foreground stars moving at negative velocity would be found on both positive and negative $l$. 

So far, APOGEE has no data available at longitudes at the expected tangent positions of the truncation radius and the possible ring. Clearly, infrared surveys like APOGEE can penetrate into the nuclear disc and gather excellent data. A satisfactory study will require a 3D extinction map and samples covering the full range of longitudes at $|l|\lesssim2$ with a consistent selection function. Thus, this work should be understood as a starting point for more comprehensive surveys of this exciting region.

\section*{Acknowledgements}
It is a pleasure to acknowledge helpful discussions with J. Binney. We thank the anonymous referee for a constructive and helpful report. We thank M. Bergemann, J.-U. Ness, M. Sormani, M. Smith, and E. Vasiliev for very helpful comments. This work was supported by the UK STFC grant ST/K00106X/1 and by the ERC under the ERC grant agreement No.~321067.

\label{lastpage}

\end{document}